\documentclass[a4paper,12pt]{article}
\usepackage{graphicx}
\usepackage{latexsym,amssymb,amsfonts,amsbsy,amsmath}
\usepackage{authblk}
    \makeatletter
    \renewcommand\section{\@startsection{section}{1}{\z@}%
                                       {-3.5ex \@plus -1ex \@minus -.2ex}%
                                       {2.3ex \@plus.2ex}%
                                       {\normalfont\normalsize\bfseries}}
                                       \renewcommand\subsection{\@startsection{subsection}{1}{\z@}%
                                       {-3.5ex \@plus -1ex \@minus -.2ex}%
                                       {2.3ex \@plus.2ex}%
                                       {\normalfont\normalsize}}
    \makeatother
\newlength{\bibitemsep}
\newlength{\bibparskip}
\let\oldthebibliography\thebibliography
\renewcommand\thebibliography[1]{%
  \oldthebibliography{#1}%
  \setlength{\parskip}{\bibitemsep}%
  \setlength{\itemsep}{\bibparskip}%
}
\renewcommand{\vec}[1]{\boldsymbol{#1}}
\def\ssspin#1#2{\boldsymbol{\sigma}_{#1}.\boldsymbol{\sigma}_{#2}}
\def\llcol#1#2{\tilde{\lambda}_{#1}.\tilde{\lambda}_{#2}}
\def\llss#1#2{\tilde{\lambda}_{#1}.\tilde{\lambda}_{#2}\,\boldsymbol{\sigma}_{#1}\boldsymbol{\sigma}_{#2}}
\begin{document}
\title{\normalsize\bf Doubly-heavy baryons, tetraquarks\\ and related topics\footnote{To appear in the proceedings of the  Workshop \emph{Double-Charm  Baryons and Dimesons}, 
                   Bled (Slovenia),  June 17-23, 2018, ed. M.~Rosina et al. }}
\author[1]{\normalsize\underline{J.-M. Richard}\footnote{\sf j-m.richard@ipnl.in2p3.fr}}
\author[2]{\normalsize A. Valcarce\footnote{\sf valcarce@usal.es}}
\author[3]{\normalsize J. Vijande\footnote{\sf javier.vijande@uv.es}}
\affil[1]{\small Universit\'e de Lyon, Institut de Physique Nucl\'eaire de Lyon,
        IN2P3-CNRS--UCBL,\authorcr  4 rue Enrico Fermi, 69622  Villeurbanne, France}
\affil[2]{\small Departamento de F\'\i sica Fundamental e IUFFyM,
         Universidad de Salamanca,\authorcr E-37008 Salamanca, Spain}
\affil[3]{\small Unidad Mixta de Investigaci\'on en Radiof\'\i sica e Instrumentaci\'on Nuclear\authorcr en Medicina (IRIMED),
          Instituto de Investigaci\'on Sanitaria La Fe (IIS-La Fe),\authorcr
            Universitat de Valencia (UV) and IFIC (UV-CSIC), Valencia, Spain}
\date{}
\maketitle
\begin{abstract}\noindent
We review the physics of doubly-heavy baryons $QQq$ and  tetraquarks $QQ\bar q\bar q$. For the latter,  the stability is reached for large enough mass ratio $M/m$, even when spin forces and color mixing are neglected. It is thus customarily claimed that $bb\bar q\bar q$ in its ground state cannot decay into $b\bar q+b\bar q$. In some model, $cc\bar u\bar d$ is shown to be stable if color mixing and spin effects are properly taken into account. It is conjectured that some $bc\bar q\bar q'$ benefits from favorable adjustments of the gluon tubes in the confinement regime. Some recent studies of  pentaquarks and hexaquarks are also summarized.
\end{abstract}
\section{Introduction}\label{se:intro}
Double-charm physics, and more generally the physics of doubly-heavy hadrons is by now rather old.  Shortly after the prediction of charm by the GIM mechanism  \cite{Glashow:1970gm},  Lee, Gaillard and Rosner \cite{Gaillard:1974mw} wrote a seminal paper anticipating many interesting properties of charmed hadrons, including  double-charm baryons, with an empirical notation which is now obsolete. As indicated in Sec.~\ref{se:tetra}, the first speculations about $QQ\bar q\bar q$ arose in 1981~\cite{Ader:1981db}, while the first detailed quark model calculation of the doubly heavy baryons $QQq$ came in 1988 \cite{Fleck:1989mb}. Since then, significant progress has been achieved, with in particular the onset of QCD sum rules and  lattice QCD, which is discussed elsewhere in these proceedings. Also, the interaction of light quarks is treated more realistically with the implementation of chiral dynamics. In the abundant literature on $QQq$ and $QQ\bar q\bar q$, there are also papers with unjustified approximations 
that do not account for the rich and subtle few-body dynamics inside these hadrons.
\section{Doubly-heavy baryons}\label{se:QQq}
Calculating $QQq$ in a given quark model is rather straightforward, and there are several interesting  studies,  e.g., \cite{Fleck:1989mb,Matrasulov:2000us,Vijande:2004at}. In the wave function of the first levels, one observes a hierarchy of the average separations, $r(QQ)\ll r(Qq)$, which can be interpreted as a spontaneous or dynamical diquark clustering. But the first excitations occur within the $QQ$ pair  and  involve a new diquark for each level.

A word about diquarks is in order. There are models where diquarks are introduced as basic constituents. They have a number of successes, and also problems, such as deciding which quarks do cluster in $q_1q_2q_3$, and explaining some  nucleon  resonances recently seen in photoproduction, which seemingly require two internal excitations. Much more questionable is the diquark picture just as an approximation, or say a lazy way of handling  the three-body problem. If the ground-state of $abc$ is searched by solving first the $ab$ problem with the potential $V_{ab}$ alone, and then the two-body problem $[ab]c$ with $V_{ac}(r)+V_{bc}(r)$ with $r$ the distance from $c$ to the center of $[ab]$, then the algebraic energy is underestimated. A simple exercise consists of comparing $V(r)$ and the average of $V(|\vec r+\vec r'|)$ when the angles of $\vec r'$ around $\vec r$ are varied. Except in the Coulomb case (Gauss theorem), one finds a non-negligible  deviation. In other words, the effective $QQ$ interaction within 
$QQq$ is influenced by the light quark. 

Though the Born-Oppenheimer approximation was invented in 1927, it has not yet reached some remote universities. Yet, if any approximation has to be made, this is probably the most interesting. The effective $QQ$ potential in a $QQq$ baryon is the analog of the quark-antiquark potential of charmonium, which itself is also a kind of Born-Oppenheimer potential: the minimal energy of the light degrees of freedom for a given $Q\bar Q$ separation.
\section{Tetraquarks with two heavy quarks}\label{se:tetra}
Estimating the tetraquark energy and structure, even in simple quark models, involves a delicate four-body problem. There is a competition between a collective compact configuration and  a breaking into two mesons. Unfortunately, this is not always very well handled  in the literature. Some authors consistently mistreated the four-body problem in other fields and in quark models. For some other authors, this is more puzzling, as they have set benchmarks of rigor for quarkonium, but became less and less rigorous as the number of constituents was increased. \textsl{Corruptio optimorum pessima}\footnote{The corruption of the best is the worst} use to say our ancestors. 

Historically, the first study of $QQ\bar q\bar q$ was made at CERN \cite{Ader:1981db}, with the observation that the system becomes bound, below the $Q\bar q+Q\bar q$ threshold, if the mass ratio $M/m$ becomes large enough. This was confirmed by Heller et al.\ \cite{Heller:1986bt,Carlson:1988hh} and Zouzou et al. \cite{Zouzou:1986qh}. The possibility of binding $QQ\bar q\bar q$ has been rediscovered in some very recent papers, which are sometimes given credit for this idea. This corresponds to the ``11th hour effect``, \textsl{So the last will be first, and the first last} (Matthew 20.16). Another sentence of Matthew's Gospel is also cited in such circumstances, in particular by the sociologist R.~Merton~\cite{1968Sci...159...56M}: \textsl{For to him who has will more be given; and from him who has not, even what he has will be taken away}. 

The binding of $QQ\bar q\bar q$  is a chromoelectric effect at start: the tetraquark benefits from the heavy-heavy attraction that is absent form the threshold.  It was also realized that chromomagnetic effects could be decisive for $cc\bar u\bar d$, with an attraction in the light sector that is absent in the threshold. A decisive progress was accomplished by Janc and Rosina \cite{Janc:2004qn}, who showed that $cc\bar u\bar d$ is stable in  a specific quark model when chromo-electric and magnetic effects are properly combined. Their result was confirmed and improved by Barnea et al.~\cite{Barnea:2006sd}. See, also, \cite{Czarnecki:2017vco}. 

There are very few rigorous results for the four-body problem, besides the ones shared with any $N$-body problem, such as the virial theorem and the scaling properties in a power-law potential. The physics of tetraquarks, however, stimulated some contributions: the improved stability when charge-conjugation symmetry $C$ is broken, and the improved stability for asymmetric potentials, as explained below. The first point should have been borrowed from atomic physics, but paradoxically, the quark physics helped to understand the transition from the positronium molecule to the hydrogen one \cite{Richard:1992cb,1993PhRvL..71.1332R}.\footnote{To be honest, a similar reasoning was already outlined in the physics of excitons \cite{1972PMag...26..143A}}. 

It is well-known that breaking a symmetry lowers the ground-state energy. For instance, going  from $H_0=p^2+x^2$ to $H_0+\lambda\,x$ lowers the first energy from $E_0=1$ to $E_0-\lambda^2/4$, and more generally, breaking parity in $H=H_\text{even}+H_\text{odd}$ gives $E<E_{even}$. But in a  few-body system, the breaking of symmetry often benefits more to the threshold than to the collective configuration and thus spoils the binding. For instance, in atomic physics, going from Ps$_2$ to $(M^+,m^+,M^-,m^-)$ makes the system unstable for $M/m\gtrsim 2.2$ \cite{PhysRevA.57.4956,1999fbpp.conf...11V}. However, when the symmetry is charge-conjugation, the symmetry breaking benefits entirely to the collective state. Let us, indeed, write the four-body Hamiltonian of the hydrogen molecule as
\begin{equation}
\begin{aligned}
H & = \frac{\vec p_1^{\,2}}{2 \, M} +
\frac{\vec p_2^{\,2}}{2 \, M} +
\frac{\vec p_3^{\,2}}{2 \, m} +
\frac{\vec p_4^{\,2}}{2 \, m} + V =H_\text{even}+H_{odd} \\
&= \left[\sum_{i}\frac{\vec p_i^{\, 2}}{2 \, \mu} \, + \, V \right] \, + \, \left(\frac{1}{4\,M} \, - \, \frac{1}{4\,m} \right)\left(
\vec p_1^{\,2} + \vec p_2^{\,2} - \vec p_3^{\,2} - \vec p_4^{\,2} \right) \, ,
\end{aligned}
\label{HCP}
\end{equation}
where $2\,\mu^{-1}=M^{-1}+m^{-1}$. The $C$-parity breaking term, $H_\text{odd}$, lowers the ground state
energy of $H$ with respect to the $C$-parity even part, $H_\text{even}$, which is simply a rescaled version of
the Hamiltonian of the positronium molecule. Since $H_\text{even}$ and $H$
have the same threshold, and since the positronium molecule is stable, the hydrogen molecule
is even more stable, and stability improves when $M/m$ increases. 
Clearly, the Coulomb character of $V$ hardly matters in this reasoning, except that if the potential is not Coulombic, $V_\text{even}$ does not always   support a bound state: in this case, stability occurs starting from a minimal value of $M/m$. 
The key assumption is that the potential does not change when the masses are modified, a property named ``flavor independence'' in QCD. 

\begin{figure}[ht!]
 \centering
 \includegraphics[width=.4\textwidth]{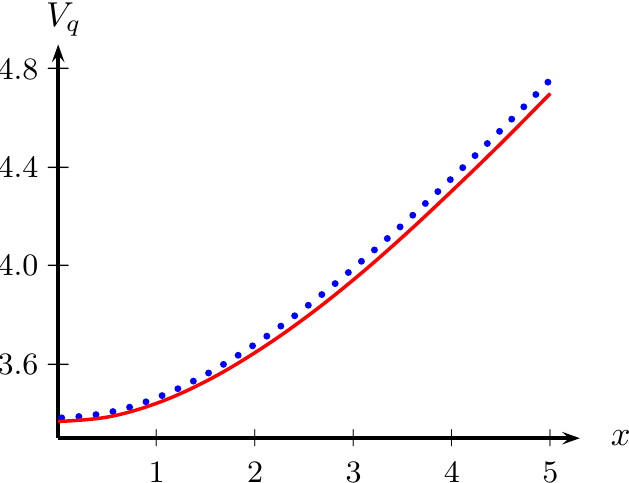}
 \caption{Comparison of the $QQ$ Born-Oppenheimer potentials in  $QQq$ (solid line) and $QQ\bar q\bar q$ (dotted line), the latter shifted by the mass difference $Qqq-\bar Q q$}
 \label{fig:BO}
\end{figure}

As ever, the Born-Oppenheimer approach is very instructive. If one restricts to color $\bar33$, the Born-Oppenheimer $QQ$ potential of $QQ\bar q\bar q$ is similar to the one of $QQq$, up to an overall constant, which can be identified as the mass difference $Qqq-\bar Qq$ from the values at zero separation. See Fig.~\ref{fig:BO}. One thus gets a microscopic derivation of the Eichten-Quigg identity (here without the spin refinements) \cite{Eichten:2017ffp}
\begin{equation}
 QQ\bar q\bar q\simeq QQq+ Qqq-\bar Q q~. 
\end{equation}
Of course, with color mixing, the mass of the tetraquark decreases with respect to the above estimate, and this can be decisive in the charm sector.

\begin{figure}[t]
\centering
\includegraphics[width=.4\columnwidth]{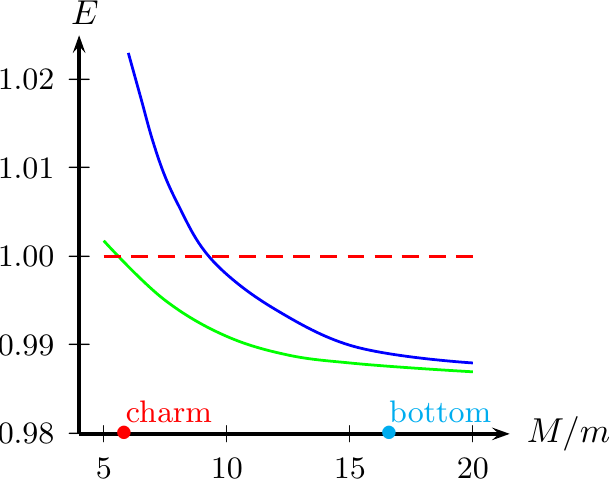}\hspace{.3cm}
\includegraphics[width=.4\columnwidth]{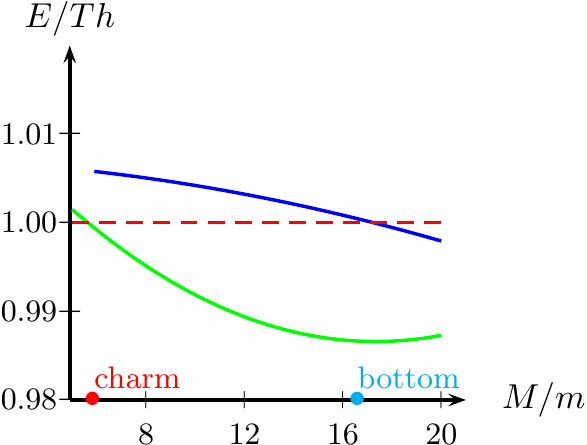}
\caption{ Effect of color-mixing (left)  and spin effects (right) on the binding of $QQ\bar u\bar d$. 
Left: the tetraquark energy is calculated with only the color $\bar33$ configurations (upper  curve) and with 
the $6\bar6$ components (lower curve). Right: the tetraquark energy calculated without (upper curve) or  with (lower curve) the chromomagnetic term. The threshold is indicated as a dashed line.}
\label{fig:col-spin}
\end{figure}

A conservative conclusion, in most studies, is that only $bb\bar q\bar q$ is stable. This is indeed the case if spin corrections and color mixing are neglected. With proper inclusion of both color $[QQ][\bar q\bar q]= \bar 33$ and $6\bar 6$ states, and spin effects, one gains some binding in the $cc\bar u\bar d$ case. This is shown in Fig.~\ref{fig:col-spin}. 

\begin{figure}[!ht]
\centering
\raisebox{.8cm}{\includegraphics[width=.15\columnwidth]{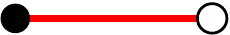}}\hspace{.6cm}
\includegraphics[width=.15\columnwidth]{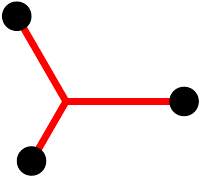}\hspace{.6cm}
\includegraphics[width=.15\columnwidth]{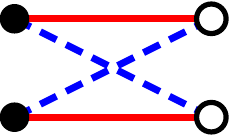}\hspace{.6cm}
\includegraphics[width=.15\columnwidth]{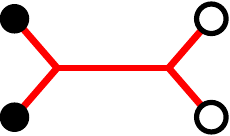}
\caption{String configurations. From left to right: mesons, baryons, flip-flop and connected double $Y$ for tetraquarks}
\label{fig:strings}
\end{figure}

Another effect could benefit to $bc\bar q\bar q$ states. A typical quark model potential reads
\begin{equation}\label{eq:pot}
 V=-\frac{3}{16}\sum_{i<j}\llcol{i}{j}\left[-\frac{a}{r_{ij}}+b\,r_{ij}+\frac{\ssspin{i}{j}}{m_i\,m_j}\,v_{ss}(r_{ij})\right]~.
\end{equation}
The linear part in \eqref{eq:pot} is interpreted as a string linking the quark to the antiquark. For baryons, it becomes the so-called $Y$-shape confinement: the three strings join at the Fermat-Torricelli point, to minimize the cumulated length. For  a system of two quarks and two antiquarks, a generalization consists of a minimization over the flip-flop and connected double $Y$ arrangements, shown in Fig.~\ref{fig:strings}. 
The changes with respect to the additive model are minor for baryons, but for  tetraquarks, the good surprise is that the flip-flop gives more attraction \cite{Vijande:2011im,Vijande:2013qr}, provided the system can evolve freely from one configuration to another one. For identical quarks and/or antiquarks, this is restricted by the Pauli principle. Thus multiquarks with non-identical quarks benefit much better from the string-mediated  dynamics.  In the future, a comparison of $bb\bar q\bar q$, $cc\bar q\bar q$ and $bc\bar q\bar q$ could probe this effect. 

Before leaving the tetraquark sector, let us discuss the all-heavy case $QQ\bar Q\bar Q$.  It is sometimes claimed to be bound below the $Q\bar Q+Q\bar Q$ threshold, but this is not the case in standard quark models, at least when treated correctly. One may wonder why Ps$_2$ is demonstrated to be bound \cite{PhysRev.71.493}, and $QQ\bar Q\bar Q$ found unstable in a simple chromoelectric model.  Let us consider the generic four-body Hamiltonian
\begin{equation}
 H_4=\sum_i \vec p_i^2+\sum_{i<j} g_{ij}\,V(r_{ij})~,
\end{equation}
where $V$ is attractive (or dominantly attractive) and $\sum g_{ij}=2$. For instance, Ps$_2$ corresponds to $V=-1/r$ and $g_{ij}=\{-1,-1,+1,+1,,+1\}$, a tetraquark with color $\bar 33$  to $\{1/2,1/2,1/4,1/4,1/4,1/4\}$,  a tetraquark with color $6\bar6$ to $\{-1/4,-1/4,5/8,5/8,5/8,5/8\}$, and the threshold to $\{1,1,0,0,0,0\}$ with a suitable renumbering. The variational principle immediately tells us that the symmetric set of strengths $g_{ij}=1/3\ \forall i<j$ maximizes the energy, and that increasing the asymmetry of the $g_{ij}$ distribution decreases the energy.  The $\chi^2$ of the distribution is larger for Ps$_2$ than for the threshold, and this explains why Ps$_2$ is stable (of course, this is not written exactly in this manner in the textbooks on quantum chemistry!). On the other hand, both color $\bar33$ and $6\bar6$ states have a $\chi^2$ smaller than the threshold and thus cannot bind.\footnote{To be more precise, if one considers a distribution $\{g_{ij}\}=\{1/3+2\,\lambda, 1/3+2\,\lambda,,1/3-\lambda,1/3-\lambda,1/3-\lambda,1/3-\lambda\}$, $E(\lambda_2)<E(\lambda_1)$ is rigorous if $\lambda_2<\lambda_1<0$ or $\lambda_2>\lambda_1>0$, while it is only most plausible if $|\lambda_2|>|\lambda_1|$ with $\lambda_1\,\lambda_2<0$, as $E(\lambda)$ is nearly parabolic as a function of $\lambda$.}\@ Numerical calculations show that instability remains when the mixing of color states is accounted for. 

\section{Pentaquarks and hexaquarks}
Other configurations are regularly revisited, with the hope to predict new stable or metastable multiquarks. 

In the pentaquark sector, the $\bar Q qqqq$ systems have been revisited. In 1987, it was shown that in the limit where $Q$ is infinitely heavy, and $qqqq=uuds,\, ddus$ or $ssud$ in the SU(3)$_\text{F}$ limit, with the assumptions that the strength of the chromomagnetic term is the same as for ordinary baryons, this state is bound by about $150\,$MeV below the $\bar Q q+qqq$ threshold. This pentaquark was searched for in an experiment at Fermilab \cite{Aitala:1997ja,Aitala:1999ij}, which turned out inconclusive. The non-strange variant was studied at HERA \cite{Aktas:2004qf,Chekanov:2004qm}.  

More precisely, if $A= \langle -\sum \llss{i}{j}\rangle$ is the expectation value of the chromomagnetic operator for $N$ or $\Lambda$, then $\bar Q qqqq$ gets $2\,A$ in the most favorable case. In further studies, it was noticed that as in the case of the famous $H=uuddss$, the multiquark wave function is more dilute than the baryon wave function. This reduces the effectiveness of the chromomagnetic interaction. This is confirmed in our recent study. 

Two contributions deal with the hidden-charm states, say $\bar Q Qqqq$, which have been much studied after the discovery of the so-called LHCb pentaquarks \cite{Aaij:2015tga}.  First, it is found that within a standard quark model of the type \eqref{eq:pot}, some states are likely below the threshold \cite{Richard:2017una}. This means that new pentaquarks perhaps await discovery, with different quantum numbers. 

Another study deals with the states in the continuum. In the early days of the quark model applied to the multiquark sector bound-state techniques were innocently applied to resonances, with the belief that if a state is found, say, 100\,MeV above the threshold using a crude one-Gaussian variational wave function, a resonance is predicted at about this energy!  The method of real scaling was applied recently to $\bar c c uud$ \cite{Hiyama:2018ukv}, using a standard quark model. It is found that one can separate clearly states that just mimic the continuum from genuine resonances. This is very encouraging, though the candidates for $(3/2)^-$ or $(5/2)^-$ are significantly higher that the LHCb pentaquarks. 

In the hexaquark sector, there is a continuous effort from many authors. Our contribution deals with $QQqqqq$, that looks at first very promising, as it combines the chromoelectric attraction of the $QQ$ pair, which acts in the threshold $QQq+qqq$, but not in $Qqq+Qqq$, and the chromomagnetic attraction which is more favorable in the latter than in the former threshold. Moreover, for $qqqq=uuds,\, ddsu$ or $ssud$, the same coherence as in the $\bar Q qqqq$ pentaquark could help. However, our study shows that the various effects hardly act together, as each of them requires a specific color-spin configuration. 
\section{Outlook}
The physics of multiquark is of primordial importance for hadron spectroscopy. The constituent models, however simple, are a good guidance before considering more ambitious theories. They require some care, but benefit of the know-how accumulated in other branches of few-body physics. Some further developments are required for describing states in the continuum. The method of real scaling looks rather promising, but might be challenged by other schemes. The coupling of channels also reveals interesting features and offers a somewhat complementary point of view \cite{2009FBS....45...99V,Garcilazo:2018san}. The transition from short-range dynamics in terms of quarks, to a long-range hadron-hadron dynamics is probably the key  to describe most of the states. 
\subsection*{Acknowledgments} JMR would like to thank M.~Rosina and his colleagues for the friendly and stimulating atmosphere of the Bled meeting, and also E.~Hiyama, M.~Oka and A.~Hosaka for a fruitlful collaboration about states in the continuum.
\begin{small}

\end{small}
\end{document}